\documentclass[aps,prb,superscriptaddress]{revtex4}
\usepackage{graphicx}
\usepackage{amsfonts}
\usepackage{amsmath}
\usepackage{amssymb}

\begin{document}

\title{Josephson current in a superconductor -- ferromagnet -- superconductor junction with in-plane ferromagnetic domains}
\author{B.\ Crouzy}
\affiliation{Institute for Theoretical Physics, \'Ecole Polytechnique F\'ed\'erale de
Lausanne (EPFL), CH-1015 Lausanne, Switzerland}
\author{S.\ Tollis}
\affiliation{Institute for Theoretical Physics, \'Ecole Polytechnique F\'ed\'erale de
Lausanne (EPFL), CH-1015 Lausanne, Switzerland}
\author{D.\ A.\ Ivanov}
\affiliation{Institute for Theoretical Physics, \'Ecole Polytechnique F\'ed\'erale de
Lausanne (EPFL), CH-1015 Lausanne, Switzerland}
\date{\today}

\begin{abstract}
We study a diffusive superconductor--ferromagnet--superconductor (SFS) junction 
with in-plane ferromagnetic domains. Close to the superconducting transition temperature,
we describe the proximity effect in the junction with the linearized Usadel equations.
We find that properties of such a junction 
depend on the size of the domains relative to the magnetic coherence length.
In the case of large domains, the junction exhibits transitions to the $\pi$ state,
similarly to a single-domain SFS junction. In the case of small domains, the magnetization
effectively averages out, and the junction is always in the zero state, similarly to
a superconductor--normal metal--superconductor (SNS) junction. In both those regimes,
the influence of domain walls may be approximately described as an effective spin-flip
scattering. We also study the inhomogeneous distribution of the local current density
in the junction. Close to the $0$--$\pi$ transitions, the directions of the critical
current may be opposite in the vicinity of the domain wall and in the middle of the
domains.
\end{abstract}

\maketitle

\section{Introduction}

Substantial progress has been made recently in understanding the physical properties of 
nanoscopic layered structures composed of superconducting (S) and ferromagnetic (F) materials 
(for a review, see Refs.\ \onlinecite{Rev,revbve}). In such systems, the coexistence of
superconducting and magnetic correlations  may lead to a variety of
interesting physical effects.
The exchange splitting of the Fermi level \cite{demler} in the
ferromagnet breaks the spin degeneracy, and Andreev-reflected Cooper pairs
acquire a finite momentum, which produces oscillations of the Cooper-pair wave
function. By tuning the geometric and electronic parameters, one can realize SFS junctions 
in which the superconducting order parameter is
of opposite sign in the two S electrodes.\cite{bula} 
Such a $\pi $ state manifests itself in a
reversal of the sign of the critical current as
the thickness of the ferromagnets is varied.\cite{Rev} The characteristic
thickness at which the first $0$--$\pi $ transition occurs is of the order
of the magnetic coherence length $\xi _{h}$. In the diffusive limit, which is realized in most 
experimentally fabricated SF heterostructures, $\xi _{h}$ is given by $\sqrt{D/h}$ 
where $D$ denotes the diffusion constant and $h$ is the ferromagnetic exchange energy. 
Therefore, the experimental observation of such $0$--$\pi $ transitions in 
nanoscale devices requires a low exchange energy $h$. This stringent condition was achieved using
weak ferromagnetic CuNi or PdNi alloys.\cite{kontos,pi1st,pi2nd,Dead} 

The physics of single-domain SFS junctions (including the effect of 
spin-flip\cite{spinfliptheo1,spinfliptheo2,spinflipexpe} and spin-orbit 
scattering\cite{faure,demler,spinorbittheo}) is now well understood. 
However, in some experiments the $0$--$\pi$ transition points may 
deviate from standard predictions\cite{Dead,shelukhin} or even be absent.\cite{koorevaar,verbanck} 
There is no consensus on the interpretation of such deviations. It
may be attributed either to the presence of a magnetically dead layer at the interface 
between the superconductor and the ferromagnet,\cite{Dead,dead2} or to a domain structure 
or inhomogeneities in the ferromagnetic layer. The domain structure crucially depends on 
the nature of the ferromagnet: strong ferromagnets consist of well-defined magnetic domains 
whose spatial extension may be reduced by the proximity effect\cite{chud1,sonin2002,chud2,younes}.
In weakly ferromagnetic alloys, on the other hand, the magnetization may fluctuate on short
length scales without forming domains.\cite{sellier} 

Theoretically, SFS junctions with inhomogeneous magnetization have been studied recently in
different setups.\cite{Bergeret,Hekking,Koshina,us2nd,fominovmultidomain,volkovmultidomain} 
However, in most works (except for Refs. \onlinecite{fominovmultidomain,volkovmultidomain}) 
only domains along the junction were studied (quasi-one-dimensional geometry), while in the
experimental realizations of SFS junctions with thin F layers the domain structure is more
likely to form in the plane of the F film.

\begin{figure}
\includegraphics[width=8.6cm]{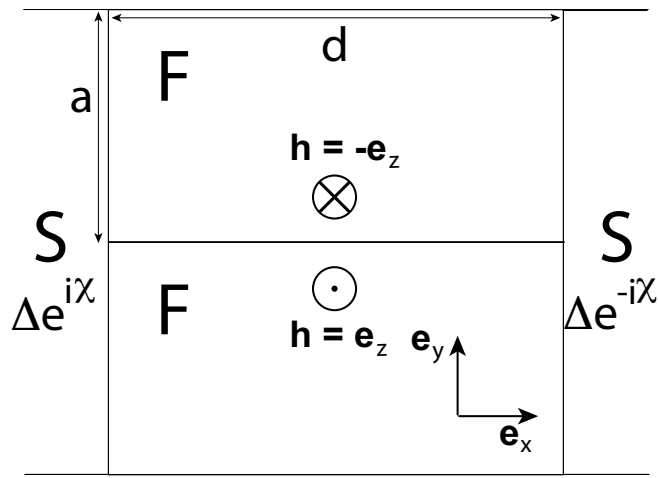} 
\caption{SFS junction with in-plane magnetic domains}
\label{setup}
\end{figure}

Motivated by the experimental progress on $\pi$ junctions, we study a model of a diffusive SFS junction
with in-plane domains (so that the domain walls are orthogonal to the S and F layers, see Fig.~1).
This geometry has been studied previously by Volkov and Anishchanka within the macroscopic approach 
of London equations.\cite{volkovmultidomain} Our model is different from the one studied in 
Ref.~\onlinecite{fominovmultidomain}: in that work, the Neel domain walls are considered, and the
junction is brought to the regime with only the long-range triplet component contributing to the
Josephson current. In our model, the domain walls are taken to be sharp, and no long-range
triplet component appears for domains with antiparallel magnetization.

The domain structure introduces an additional length scale: the domain size $a$. As one can expect, we find that
the effect of inhomogeneous magnetization depends strongly on the relative magnitude
of $a$ and $\xi_h$. In the limit of small domains, $a\ll\xi_h$, 
the exchange field effectively averages out, and the critical current of a
single nonmagnetic SNS junction is retrieved. In the opposite limit of large domains $a\gg\xi_h$,
the influence of domain walls is localized to their vicinity and produces only a small
correction to the current of a single-domain SFS junction.
Between those limits, the supercurrent shows either a damped
oscillatory behavior as a function of the junction thickness (for large domains 
$a>a_{c}\approx0.83\,\xi _{h}$), or a monotonic exponential decay (for smaller domains $a<a_{c}$). 
In the former case, the multidomain junction may be compared to a single-domain SFS trilayer 
with spin-flip scattering\cite{faure} and a renormalized exchange field, 
whereas in the latter case the junction behaves like a SNS junction with spin-flip scattering.
The effective parameters are determined analytically in both limits of small and large domains.
We also study the inhomogeneous distribution of the current density and conjecture that
at low temperatures such SFS junctions with domains may realize the intermediate $\varphi$ phase
proposed by Buzdin.\cite{phiphase}

The paper is organized as follows. In Section II we compute the superconducting
Green functions and the Josephson-current density for the multidomain SFS junction. 
Section III is devoted to the analysis of the total Josephson current. 
In Section IV we discuss the spatial distribution of the current density.
Finally, in Section V we summarize our conclusions.

\section{A model for the multidomain SFS junction}

We assume that the ferromagnetic layer is strongly disordered, and the motion of
electrons is diffusive. In this regime, the Green functions are given by the solutions
to the Usadel equations.\cite{Us} To simplify the calculations, we further assume 
that the junction is close to the superconducting critical temperature $T_{c}$.
In this case, the superconducting correlations are weak so that the Usadel equations can be
linearized, and the current-phase relation is sinusoidal%
\cite{golubov} 
\begin{equation}
J=J_{c}\sin \varphi \,,  \label{cpr}
\end{equation}%
where $\varphi=2\chi$ is the superconducting phase difference across the junction
and $J_{c}$ is the critical current. The sign of $J_{c}$ determines if  the junction
is in the ``zero'' phase or in the ``$\pi$'' phase.

In this paper, we consider a SFS junction with in-plane ferromagnetic domains 
of opposite magnetization. We introduce a coordinate system with the
F layer in the $yz$ plane (Fig.~1). The $x$ axis is directed along the 
junction, and the SF interfaces correspond to the coordinates $x=0,d$. 
The domain walls are taken to be normal to the $y$ axis. The origin of the
$y$ axis is chosen at the interface between two domains. 
The system is invariant under translation
along the $z$ axis. Our further calculations will be equally applicable
to either the system with
two domains of width $a$ (see Fig.~1) or the $2a$-periodic multidomain case 
(the same setup periodically repeated in the $y$ direction).

The (nonlinear) Usadel equation in the ferromagnetic layer takes the form (we follow the
conventions used in Ref.\ \onlinecite{I+F}) 
\begin{equation}
D\nabla \left( \check{g}\nabla \check{g}\right) -\omega \left[ \hat{\tau}_{3}%
\hat{\sigma}_{0},\check{g}\right] -i\left[ \hat{\tau}_{3}\left( \mathbf{%
h\cdot \mathbf{\hat{\sigma}}}\right) ,\check{g}\right] =0.  \label{usadel}
\end{equation}%
where $D$ denotes the diffusion constant and the system of units with 
$\hbar =k_{B}=\mu _{B}=1$ is chosen. The Green function $\check{g}$ is a
matrix in the Nambu $\otimes $ spin space, $\hat{\tau}_{\alpha }$ and $\hat{%
\sigma}_{\alpha }$ denote the Pauli matrices respectively in Nambu
(particle-hole) and spin space, $\omega =\left( 2n+1\right) \pi {T}$ are the
Matsubara frequencies and $\mathbf{h}$ is the exchange field in the
ferromagnet. The Usadel equation is supplemented with the normalization
condition for the quasiclassical Green function 
\begin{equation}
\check{g}^{2}={\check{1}}=\hat{\tau}_{0}\hat{\sigma}_{0}.
\label{normalization}
\end{equation}%
For simplicity, we assume that the superconductors are much less disordered
than the ferromagnet, and then we can impose the rigid boundary conditions at
the SF interfaces,
\begin{equation}
\check{g}=\frac{1}{\sqrt{\omega ^{2}+\Delta ^{2}}}\left( 
\begin{array}{cc}
\omega  & \Delta e^{\pm {i}\chi } \\ 
-\Delta e^{\mp {i}\chi } & -\omega 
\end{array}%
\right) _{\mathrm{Nambu}}\otimes \hat{\sigma}_{0}\,,  \label{bc}
\end{equation}%
where $\Delta $ denotes the superconducting order parameter, and the
different signs refer respectively to the boundary conditions at $x=0$ and $%
x=d$. At the (transparent) interface between the two ferromagnetic domains,
we impose the continuity of the Green functions and of their derivatives.

Close to the critical temperature $T_{c}$, we linearize the Usadel
equations (\ref{usadel}), (\ref{normalization}) around the solution for the normal metal state 
$\check{g}=\hat{\tau}_{3}\hat{\sigma}_{0}\mathrm{sgn}(\omega)$. The linearized Green function then takes the
form 
\begin{equation}
\check{g}=\left( 
\begin{array}{cc}
\sigma _{0}\mathrm{sgn}(\omega ) & f_{\alpha }\sigma ^{\alpha } \\ 
-f_{\alpha }^{\dagger }\sigma ^{\alpha } & -\sigma _{0}\mathrm{sgn}(\omega )%
\end{array}%
\right) ,
\end{equation}%
where the scalar $f_{0}$ (respectively $f_{0}^{\dagger }$) and vector $%
\mathbf{f}$ (respectively $\mathbf{{f}^{\dagger }}$) components of the
anomalous Green functions obey the linear equations 
\begin{equation}
\left(\frac{\partial ^{2}}{\partial x^{2}}+\frac{\partial
^{2}}{\partial y^{2}}\right)f_{\pm }^{1(\dagger )}- \lambda _{\pm }
^{2}f_{\pm }^{1(\dagger )}=0,  \label{lin}
\end{equation}%
with 
\begin{equation}
\lambda _{\pm }=\left[ 2\frac{\left\vert \omega \right\vert \mp ih\mathrm{sgn%
}(\omega )}{D}\right] ^{1/2}.
\end{equation}%
The projections of the anomalous Green function along the direction of the
exchange field (\textquotedblleft parallel\textquotedblright\ components)
are defined as $f_{\pm }^{(\dagger )}(x,y)=f_{0}^{(\dagger )}\pm \mathbf{f}%
^{(\dagger )}\cdot \mathbf{e}_{z}$ (we assume that the ferromagnetic 
exchange field ${\mathbf h}$ is aligned in the direction $\mathbf{e}_{z}$,
see Fig.~1). Note that there is no perpendicular
("long-range triplet"\cite{Bergeret-long-range}) component of the vector part of the Green 
function, since the
magnetizations of the domains are collinear. We have used
the invariance under translation along the $z$ direction. The superscript 1 refers 
to domains with field along $\mathbf{e_{z}}$ and in the following we will use 
the superscript 2 for domains with the field along $-\mathbf{e_{z}}$. 
Similar equations hold for $f_{\pm }^{2(\dagger )}$ with $\lambda_{\pm}\leftrightarrow\lambda_{\mp}$.

It is convenient to write the solutions to those equations in the form 
\begin{equation}
f_{\pm }^{(\dagger )1,2}(x,y)=f_{\pm \mathrm{Bulk}}^{(\dagger
)1,2}(x)+\delta _{\pm }^{(\dagger )1,2}(x,y)  \label{greenfunction}
\end{equation}%
where $f_{\pm \mathrm{Bulk}}^{1,2}$ are the solutions of Eq. (\ref{lin}) for a 
single-domain SFS junction with the magnetization along $\mathbf{e_{z}}$ 
(respectively $-\mathbf{e_{z}}$). Since the equations (\ref%
{lin}) are linear, the correction $\delta (x,y)$ is also a solution to the same
equations with the boundary conditions 
\begin{eqnarray}
&&\partial _{y}\delta _{\pm }^{1,2}(x,y=\mp{a}) =0,  \label{vacuum} \\
&&\delta _{\pm }^{1,2}(x=\{0,d\},y) =0,  \label{SFinterface} \\
&&\delta _{\pm }^{1}(x,y=0)- \delta _{\pm }^{2}(x,y =0)= \Delta {f}%
_{\pm \mathrm{Bulk}}(x),  \label{middlex} \\
&&\left[ \partial _{y}\right] \delta _{\pm }^{1}(x,y=0)-\left[ \partial _{y}%
\right] \delta _{\pm }^{2}(x,y =0)=0.  \label{middley}
\end{eqnarray}%
Here $\Delta {f}_{\pm \mathrm{Bulk}}=f_{\pm \mathrm{Bulk}}^{2}(x)-f_{\pm 
\mathrm{Bulk}}^{1}(x)$ is the difference of the bulk Green functions in the two
domains. For the two-domain junction, the first condition imposes zero
current at the interface with vacuum, the second condition ensures the
continuity of the Green functions at the SF interfaces. Finally, the last two
conditions reflect the continuity of the Green function and its derivatives
at the interface between the two domains. It can be easily shown from
symmetry considerations that this set of boundary conditions can also be
applied to a periodic multidomain SFS junction with domains of width $2a$.

The condition (\ref{SFinterface}) allows us to express $\delta _{\pm }^{1,2}$ in the
form of the Fourier series 
\begin{equation}
\delta _{\pm }^{1,2}=\sum_{n=1}^{\infty }\sin \left( \frac{\pi {n}}{d}%
x\right) A_{n\pm }^{1,2}(y).  \label{series}
\end{equation}%
For each $n$ we solve 
\begin{equation}
\partial _{y}^{2}A_{n\pm }^{1}=\gamma _{n\pm }^{2}A_{n\pm }^{1}  \label{coeff}
\end{equation}%
with 
\begin{equation}
\gamma_{n\pm }=\sqrt{(\frac{\pi {n}}{d})^2+\lambda _{\pm }^2}\, .
\end{equation} 
To obtain
the equation for $A_{n\pm }^{2}$ one needs to substitute $\gamma _{n\pm
}\leftrightarrow \gamma _{n\mp }$. We can solve those equations for each
Fourier component $n$  with the boundary conditions provided by 
(\ref{vacuum}), (\ref{middlex}) and (\ref{middley}). The solution is given by
\begin{equation}
\delta _{\pm }^{1}=\frac{\Delta}{|\omega|} \sum_{n=1}^{\infty }%
\sin \left( \frac{\pi {n}x}{d}\right) \frac{2\pi {n}}{d^{2}}\frac{\cosh
\gamma _{n\pm }(y+a)}{\cosh \gamma _{n\pm }a}\frac{\gamma _{n\mp }\tanh
\gamma _{n\mp }a}{\gamma _{n\mp }\tanh \gamma _{n\mp }a+ \gamma _{n\pm }\tanh \gamma _{n\pm }a}\left( \frac{1}{%
\gamma _{\mp }^{2}}-\frac{1}{\gamma _{\pm }^{2}}\right) \left( e^{i\chi
}-\left( -1\right) ^{n}e^{-i\chi }\right)\, .  \label{delta1}
\end{equation}%
In the second domain, the correction $\delta _{\pm }^{2}$ is given by the same formula with
the replacement of $y,\gamma_{\pm}$ by $-y,\gamma_{\mp}$. \ The bulk Green functions are given by%
\cite{us2nd} 
\begin{equation}
f_{\pm \mathrm{Bulk}}^{1}=\frac{\Delta}{|\omega|}\left[ \frac{\sinh
\lambda _{\pm }x}{\sinh \lambda _{\pm }d}e^{-i\chi }+\frac{\sinh \lambda
_{\pm }(d-x)}{\sinh \lambda _{\pm }d}e^{i\chi }\right],\label{bulk}
\end{equation}%
and $f_{\pm \mathrm{Bulk}}^{2}=f_{\mp  \mathrm{Bulk}}^{1}$. 
Finally, note that $f_{\mathrm{%
Bulk}}^{\dagger }$ and $\delta ^{\dagger }$ are given by the same expressions
(\ref{delta1}), (\ref{bulk}) with the replacement of $\chi$ by $-\chi $. 

The last step will be to compute the Josephson current density using the
formula\cite{revbve} 
\begin{equation}
\mathbf{J}=ieN(0)D\pi T\sum_{\omega =-\infty }^{\infty }\frac{1}{2}\mathrm{Tr%
}\left( \hat{\tau}_{3}\hat{\sigma}_{0}\check{g}\mathbf{\nabla }\check{g}%
\right) ,
\end{equation}%
where $N(0)$ is the density of states in the normal metal phase (per one
spin projection) and the trace has to be taken over the Nambu and spin indices.
The current density can be explicitly
rewritten for the linearized $\check{g}$ 
\begin{equation}
\mathbf{J}=-ieN(0)D\pi T\,\sum_{\omega =-\infty }^{\infty }\left[ \sum_{\sigma
=\pm }\frac{1}{2}(f_{\sigma }\mathbf{\nabla }{f}_{\sigma }^{\dagger
}-f_{\sigma }^{\dagger }\mathbf{\nabla }{f}_{\sigma })\right].
\end{equation}%
The symmetry of translation along the $z$ direction implies that the current
remains in the $xy$ plane. Using the expression for the Green functions (\ref%
{delta1}) and (\ref{bulk}), we can obtain a general expression for the
current density (which is too cumbersome to be reproduced here). 
This expression involves two contributions. The first one is produced exclusively
by the bulk Green functions (\ref{bulk}) and corresponds to a homogeneous ferromagnetic interlayer.
The second contribution is due to the correction (\ref{delta1}) and reflects
the influence of the domain structure. The current resulting from this contribution 
is not uniform in space.  The characteristic decay scale of this correction as a function 
of the distance from the domain interface is given by 
$\Re\left(\frac{1}{\gamma _{n\pm }}\right)\sim \min (\xi _{T},\xi _{h},d)$, 
where $\xi _{T}=\sqrt{D/2\pi {T_{c}}}$ and $\xi _{h}=\sqrt{D/h}$ are the
thermal and magnetic coherence lengths, respectively. 
Far from the interface between the domains 
($y\gg \min (\xi_{T},\xi _{h},d)$), the correction (\ref{delta1}) 
vanishes and we recover locally the single-domain SFS current. Thus
we expect the properties of the junction to be very different in the two
limits of small [$a\ll\min (\xi _{T},\xi _{h},d)$] and large 
[$a\gg \min(\xi _{T},\xi_{h},d)$] domains.

\section{Critical current}

Experimentally, in SFS hybrid junctions, the measurable quantity is the
total current flowing through the junction, 
that is along the $x$-axis. Since $\mathbf{%
\nabla }\cdot \mathbf{J}=0$, the total current is conserved along the $x$
direction. We can therefore compute it at $x=0$, and we find 
\begin{equation}  \label{integrated}
\frac{J_{c}}{I_{0}} 
=\Re \left[ \sum\limits_{\omega >0}\frac{\Delta ^{2}}{\omega ^{2}}%
\frac{\lambda _{+}d}{\sinh \lambda _{+}d}\right] +\frac{16\pi ^{2}}{ad^{2}
\xi_{h}^{4}}\sum\limits_{\omega >0}\frac{\Delta
^{2}}{\omega ^{2}}\sum\limits_{n=1}^{\infty }\left[ \frac{(-1) ^{n-1} n ^{2}}{
(\gamma _{n+} \gamma _{n-}) ^{3}} \frac{1}{\gamma _{n-} \coth \gamma _{n+} a + \gamma _{n+} \coth \gamma _{n-} a} \right]  
\end{equation}
with 
\begin{equation}
I_{0}=\frac{4eN(0)DS\pi T}{d},
\end{equation}
and $S$ the area of the junction. The first term is the critical current for a single-domain SFS 
junction with a damped oscillatory dependence on the F-layer thickness (for a review, 
see Refs. \onlinecite{Rev}, \onlinecite{buzdin82} and \onlinecite{buzdin91}). It can be 
either positive (zero state of the junction) or negative ($\pi $ state). The second term 
reflects the influence of the domain structure. The critical current (\ref{integrated}) depends
on the three dimensionless parameters: $a/\xi_h$, $d/\xi_h$, and $\xi_T/\xi_h$.
For some values of the parameters, the critical current (\ref{integrated})
computed numerically is plotted in Fig.~2. Depending on the values of the
parameters, it shows either an exponential decay or an exponential decay
with oscillations, as a function of $d$.

\begin{figure}
\includegraphics[width=8.6cm]{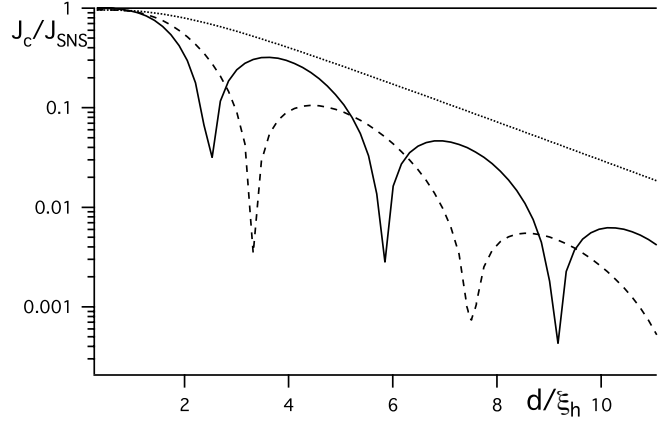} 
\caption{Critical current $J_{c}/J_{SNS}$ vs. junction length $d/\xi_{h}$ for $a=0.6\cdot\xi_{h}$ (dotted line), $1.6\cdot\xi_{h}$ (dashed line) and $\infty$ (solid line). We take $\frac{h}{T}=100$. }
\label{integrated_current}
\end{figure}

Note that in most experimental situations $\xi_T\gg\xi_h$, because the 
ferromagnetic exchange energy exceeds by far the superconducting critical 
temperature. In the following we will refer to this situation as the 
high-field limit. In this limit, the summation over $\omega$ in 
Eq.~(\ref{integrated}) can be performed analytically 
[$\sum_{\omega>0} \Delta^2/\omega^2 = \Delta^2/(8 T^2)$], and the deviation $\delta J_c$
from the critical current of a single-domain SFS junction 
[the second term in Eq.~(\ref{integrated})] is expressed in terms of the reduced variables 
$n^{*}=\frac{d\sqrt{2}}{\pi \xi_{h}}$ and $a^{*}=\frac{\pi a}{d}$:
\begin{equation}  \label{courantsimple}
\frac{\delta J_{c}}{I_{0}} =-\frac{\Delta^{2}}{2T^{2}}\frac{n^{*4}}{a^{*}}\sum\limits_{n=1}^{\infty } \frac{(-1)^{n}n^{2}}{(n^{4}+n^{*4})^{3/2}\Re \left[
\sqrt{n^{2}+in^{*2}}\coth \left(a^{*}\sqrt{n^{2}-in^{*2}} \right) \right]}.
\end{equation}
In the limit of large $d$, the asymptotic behavior of this expression
may be estimated as an integral (in the variable $z=n/n^*$)
\begin{equation}  \label{oscillating-integral}
\frac{\delta J_{c}}{I_{0}} =-\frac{\Delta^{2}}{2T^{2} a^*}
\int\limits_{-\infty}^{\infty } dz\, \frac{e^{i\pi n^* z} z^{2}}
{(z^{4}+1)^{3/2}\left[
\sqrt{z^{2}+i}\coth \left(a^{*}n^* \sqrt{z^{2}-i}\right) +
\sqrt{z^{2}-i}\coth \left(a^{*}n^* \sqrt{z^{2}+i}\right)
 \right]}\, ,
\end{equation}
which is, in turn, determined to the exponential precision by the 
singularities of the integrand in
the complex plane. Remarkably, the contribution
from the poles at $(\pm i)^{1/2}$ cancels 
exactly the first term (single-domain contribution) of Eq.~(\ref{integrated}). 
For sufficiently large $d$, to the exponential precision, the critical
current is then given by
\begin{equation}
J_c \propto e^{-\lambda d}\, , \qquad
\lambda=-\frac{i \sqrt{2} z_0}{\xi_h}\, ,
\label{lambda-definition}
\end{equation}
where $z_0$ is the singularity of the integrand with the smallest positive
imaginary part. Note that $z_0$ is now a function of one dimensionless 
parameter $\alpha= a^* n^*=\sqrt{2} a/\xi_h$.

By analogy with a single-domain SFS junction with spin-flip scattering,
the real and imaginary parts of $\lambda^2$ may be interpreted as an effective
magnetic field and an effective spin-flip rate 
\footnote{We define the spin-flip scattering rate by 
$\Gamma_{sf}=1/(2\tau_{sf})$, as in Refs.~\onlinecite{Be,Crouzy,I+F}},
\begin{equation}
\lambda^2= - \frac{2i}{\left[\xi_h^{\rm (eff)}\right]^2} + 
\frac{4 \Gamma_{sf}^{\rm (eff)}}{D}\, , \qquad 
\xi_h^{\rm (eff)} = \sqrt\frac{D}{h^{\rm (eff)}} \, .
\end{equation}
Therefore the effective field and spin-flip rate can be found as
\begin{equation}
h^{\rm (eff)} = h \Im (z_0^2) \, , \qquad 
\Gamma_{sf}^{\rm (eff)} = -\frac{h}{2} \Re (z_0^2)\, .
\end{equation}

\begin{figure}
\includegraphics[width=8.6cm]{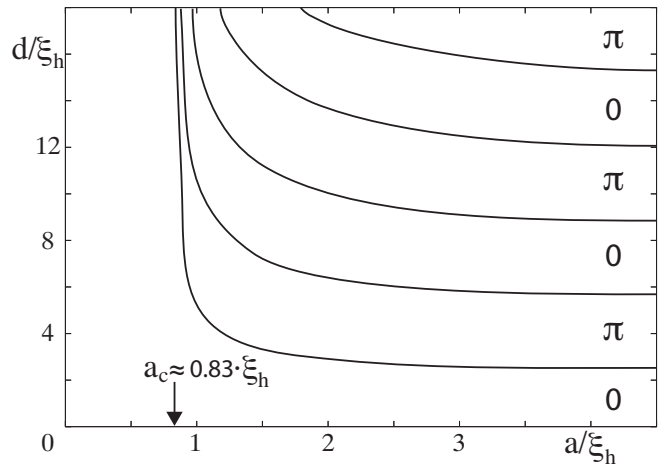} 
\caption{Phase diagram of the junction in the high-exchange-field limit. 
Here $a$ represents the width of the domains and $d$ is 
the length of the junction.}
\label{phase_diagram}
\end{figure}

In the following, we discuss the limits of large and small domain sizes.

\subsection{Limit of large domains : $a\gg \xi_{h}$}

We consider the limit of large domains, $a\gg \xi_{h}$, with the assumption
of the strong exchange field, $\xi_T\gg\xi_h$. In this regime,
the damped oscillations of the critical current at large $d$ are
determined by the solutions to the equation
\begin{equation}
\sqrt{z^2+i} \coth (\alpha \sqrt{z^2-i}) + 
\sqrt{z^2-i} \coth (\alpha \sqrt{z^2+i}) =0
\label{z-equation}
\end{equation}
with the smallest positive imaginary part. At $\alpha=\sqrt{2} a/\xi_h \gg 1$,
one of the arguments of $\coth(\alpha\sqrt{z^2\pm 1})$ must be close
to $\pm i\pi/2$. Expanding around this point, we obtain
$z_0^2= i - \frac{\pi^2}{4\alpha^2} + \frac{(1-i)\pi^2}{4\alpha^3} + \dots$
This translates into the reduced effective field
\begin{equation}
h^{\rm (eff)} \approx h \left[ 1- \frac{\pi^2}{8\sqrt{2}} 
\left(\frac{\xi_h}{a}\right)^3\right]
\end{equation}
and the effective spin-flip rate
\begin{equation}
\Gamma_{sf}^{\rm (eff)} \approx \frac{\pi^2}{16} 
\left( \frac{\xi_h}{a}\right)^2 h = \frac{\pi^2 D}{16 a^2} \, .
\end{equation}
Thus, to the leading order in $(\xi_h/a)$, the effect of domain walls
reduces to an effective spin-flip rate, which increases the period
of $0$--$\pi$ transitions as a function of $d$ and simultaneously decreases
the overall decay length of the critical current 
(see Fig.~2, dashed line, for an illustration).

\subsection{\protect\bigskip Limit of small domains : $a\ll \protect\xi %
_{h},d,\protect\xi _{T}$}

In the limit of small domains $a\ll \xi _{h},d,\xi _{T}$, we
can calculate a perturbative correction to the critical current
by expanding (\ref{integrated}) in $a$. To the lowest order in $a$,
we obtain (without assuming the high-field limit),
\begin{eqnarray}
\frac{J_{c}}{I_{0}}=\frac{J_{SNS}}{I_{0}}-\frac{2a^{2}d^{2}}{3\xi _{h}^{4}}%
\sum\limits_{\omega >0}\frac{\Delta ^{2}}{\omega ^{2}}\left[ 
\frac{\lambda _{0}d\cosh \lambda _{0}d-\sinh
\lambda _{0}d}{\lambda _{0}d\sinh
^{2}\lambda _{0}d}\right],
\end{eqnarray}
where $\lambda _{0}^{2}=\frac{\lambda _{+}^{2}+\lambda _{-}^{2}}{2}=\frac{%
2\left\vert \omega \right\vert }{D}$ does not contain the exchange energy $h$,
and $J_{SNS}=J_{c}(h=0)$. 
This expression reveals that in the limit $a\to 0$ the
multidomain SFS junction behaves like a SNS junction: the exchange field is
averaged out when the domain width is small. Note also that the correction 
arising from a finite domain width is always negative: the amplitude of 
the current is decreased compared to the SNS case.

A more accurate approximation may be obtained in the high-field
limit $\xi_T\gg\xi_h$ by the asymptotic estimate of the oscillating
sum described earlier in this Section. To the second order in $a$,
the solution to the equation (\ref{z-equation}) is given by
$z_0=-\frac{\alpha^2}{3}$, which translates into
\begin{equation}
h^{\rm (eff)} =0\, , \qquad
\Gamma_{sf}^{\rm (eff)} \approx \frac{1}{3} \left(\frac{a}{\xi_h}\right)^2 h
=\frac{h^2 a^2}{3 D} 
\end{equation}
This expression for $\Gamma_{sf}^{\rm (eff)}$ agrees with the general
estimate for the effective spin-flip rate obtained by 
Ivanov and Fominov\cite{I+F} for SF structures with 
inhomogeneous magnetization.

Note that for sufficiently small $a$, the equation (\ref{z-equation}) has
a solution with real $z_0^2$ corresponding to a pure decay (without
oscillations) of the critical current. The dependence of the
critical current on $d$ is then purely decaying, without $0$--$\pi$ oscillations
(Fig.~2, dotted line).

\begin{figure}
\includegraphics[width=8.6cm]{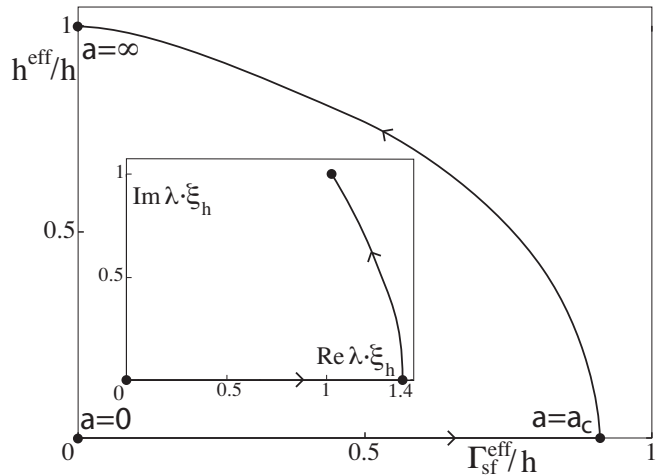} 
\caption{The effective spin-flip scattering rate $\Gamma_{sf}^{\rm (eff)}$ and the effective 
exchange field $h^{\rm (eff)}$. The curve starts at $a=0$ and ends at $a=\infty$. 
The inset shows the real part (decay length) and 
the imaginary part (rate of oscillations) of $\lambda$ in Eq.~(\ref{lambda-definition}).}
\label{effective}
\end{figure}

\subsection{$0$--$\pi$ phase diagram}

Between the two regimes of small and large domains, there is a phase
transition as a function of $a/\xi_h$ corresponding to a bifurcation
of the real solution $z_0^2$ to Eq.~(\ref{z-equation})
at smaller $a$ to complex solutions at larger $a$.
For $a/\xi_h$ smaller than the critical value, the critical current
decays as a function of $d$ without oscillations (always in the $0$ phase).
For $a/\xi_h$ larger than the critical value, the dependence on $d$ is
damped oscillatory, qualitatively similar to a single-domain SFS junction.

Numerically, we find the critical value $a_{c}/\xi_h \approx 0.83$. The full
$0$--$\pi$ phase diagram in the high-field limit is plotted in Fig.~3. Periodic
$0$--$\pi$ transitions (as a function of $d$) above $a_{c}/\xi_h$ and zero
phase below $a_{c}/\xi_h$ illustrate our discussion. The
absence of the $0$--$\pi$ transitions in the case of small domains may
explain why in some experimental SFS junctions the $\pi$ state is absent.\cite{koorevaar,verbanck}

For completeness, in Fig.~4 we also plot the locus of solutions $z_0^2$ to 
Eq.~(\ref{z-equation}) in the complex plane for all values of $\alpha$ (in the
units $\Gamma_{sf}^{\rm (eff)}/h$ and $h^{\rm (eff)}/h$). The corresponding
real and imaginary parts of $\lambda$ determining the $d$ dependence of the critical
current (\ref{lambda-definition}) are plotted in the inset.

\section{Local current density}

\begin{figure}
\includegraphics[width=8.6cm]{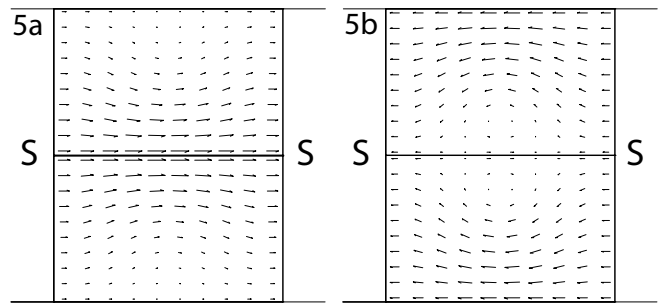} 
\caption{Josephson-current density in the two-domain SFS junction in the
zero and $\pi$ phases.
The domain size is $a=1.6\; \xi_h \ll \xi_T$, and $d$
is taken to be below (left) and above (right) 
the first $0$--$\pi$ transition.}
\label{current_lines}
\end{figure}

Since the system does not have a translational symmetry along the
$y$ direction, the Josephson current forms a nontrivial pattern
in the $x$-$y$ plane. In Fig.~5 we present plots of the current
density (proportional to $\sin\varphi$) at two different points of the
phase diagram: in the zero phase and in the $\pi$ phase.

Those inhomogeneous patterns may be qualitatively understood on the
basis of interpreting the domain walls as producing an effective
spin-flip scattering. Different regions of the ferromagnet may be
attributed different effective spin-flip rates, depending on their
distance from the domain wall. The effective spin-flip processes
renormalize the decay coefficient $\lambda$ in (\ref{lambda-definition})
and, therefore, different parts of the junction experience $0$--$\pi$
transitions at different values of $d$. This can be clearly seen in
Fig.~6 depicting the current density near a $0$--$\pi$ transition.
While the neighborhood of the domain wall is in the $0$ phase, the
region near the free boundaries (at $y=\pm a$) are in the $\pi$ phase.
This situation resembles a model studied by 
Buzdin \textit{et al.}\cite{phiphase}: a system of alternating zero and
$\pi$ junctions. In that work, an intermediate equilibrium phase difference
was predicted, depending on the ratio between the junction widths and 
the magnetic coherence length. Even though our model cannot lead
to such a $\varphi $-junction (we consider
linearized Usadel equations and therefore obtain a purely 
sinusoidal current-phase relation with only two possible 
equilibrium phases $0$ or $\pi $), at low temperatures such a SFS system with domains could possibly produce
a $\varphi$-state.

\begin{figure}
\includegraphics[width=8.6cm]{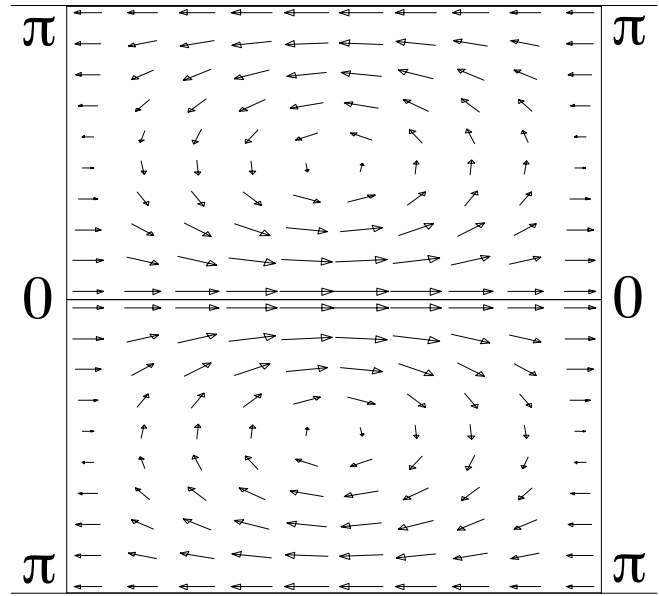} 
\caption{Current lines close to the 0--$\pi$ transition.
The domain size is $a=1.6\; \xi_h \ll \xi_T$, and $d$
is taken to be close to the first $0$--$\pi$ transition.}
\label{current_inhomogeneous}
\end{figure}

\section{Summary}

In this work we consider a Josephson SFS junction consisting of
domains with opposite magnetization
connected ``in parallel''. As a function of the junction thickness,
the critical current may exhibit either a decaying oscillating or a purely
decaying behavior, depending on the domain width. The effect of domain walls
in this geometry may be approximated as an effective spin-flip scattering,
together with a renormalization of the effective magnetic field.
This behavior is different from that in SFF'S
junctions with the domains connected
``in series'' studied in Ref.~\onlinecite{us2nd}. In that SFF'S setup,
the domain structure lead to a gradual reduction of the $\pi$ phase
(at a non-parallel configuration of the two domains), so that the
relative fraction of the zero phase increases as a function of the mismatch
in the magnetization directions. In the present work, however, we do not
consider the case of an arbitrary angle between the two magnetizations,
because of the complexity of the problem.

We expect that in a realistic geometry of domains both effects of the
spin-flip scattering and of the reduction of the $\pi$ phase take place
simultaneously, and our findings from this work and from 
Ref.~\onlinecite{us2nd} may help to qualitatively describe the $0$--$\pi$
phase diagram of real SFS junctions with inhomogeneous ferromagnets.

\begin{acknowledgments}
This work was supported by the Swiss National Foundation. We thank M. Houzet and A. Buzdin for helpful discussions.
\end{acknowledgments}
\bibliography{SFS3}
\end{document}